\documentclass[superscriptaddress,twocolumn,floatfix,nofootinbib]{revtex4-2}
\usepackage{times,fancyhdr}
\usepackage[dvips]{graphicx}
\usepackage{amsmath,amssymb,bm,dsfont}
\usepackage{color}
\usepackage{mathrsfs}
\usepackage{setspace}
\usepackage{hyperref}

\def\beq{\begin{equation}}
\def\eeq{\end{equation}}
\def\beqn{\begin{align}}
\def\eeqn{\end{align}}

\begin{document}
\title{Reply to arXiv:2111.13357 (``The Quantum Eraser Non-Paradox'')}
\author{C. Bracken}
\affiliation{Dept of Experimental Physics, Maynooth University, Maynooth, Co. Kildare, Ireland}
\affiliation{Astronomy \& Astrophysics Section, School of Cosmic Physics, Dublin Institute for Advanced Studies, Fitzwilliam Place, Dublin 2, D02 XF86}
\author{J.R. Hance}
\email{jonte.hance@bristol.ac.uk}
\affiliation{Quantum Engineering Technology Laboratories, Department of Electrical and Electronic Engineering, University of Bristol, Woodland Road, Bristol, BS8 1US, UK}
\author{S. Hossenfelder}
\affiliation{Frankfurt Institute for Advanced Studies, Ruth-Moufang-Str. 1, D-60438 Frankfurt am Main, Germany}
 
\date{\today}

\begin{abstract}
    In a recent criticism (arXiv:2111.13357) of our paper arXiv:2111.09347, Drezet argues that we have forgotten to consider superpositions of detector eigenstates. However, such superpositions do not occur in the models our paper is concerned with. We also note that no one has ever observed such detector superpositions.
\end{abstract}

\maketitle

In our recent paper \cite{Bracken2021QEParadox}, we propose a quantum mechanical experiment that leads to a temporal paradox for local-realist hidden variables models which rely on backward causation (``retrocausality''). We argue that resolving the paradox requires giving up the idea that in such models a choice can influence the past, and that it instead requires a violation of Statistical Independence without retrocausality. Throughout our paper, we state explicitly and repeatedly that we are concerned with local realist models, rather than the Bell-nonlocal/contextual models typically referred to as Standard Quantum Mechanics or any mathematically equivalent interpretations thereof.

In a local realist hidden variable model the wave-function is an ensemble of ontic states. The outcome of a measurement is always a detector eigenstate. Which of the detector eigenstates is realised -- i.e., which measurement outcome occurs -- is determined by the hidden variables. In such models, the outcome of a measurement is never a superposition of detector eigenstates. The entire purpose of such models is to make sure that the outcome of a measurement is always a detector eigenstate. We know from Bell's theorem and the observed violations of the inequality which follows from it that such models must violate Statistical Independence, hence they are either superdeterministic or retrocausal. These models are the focus of our paper.

In a recent criticism of our paper, Drezet \cite{Drezet2021CommentQEParadox} argues that we have forgotten to consider superpositions of detector eigenstates. However, such superpositions do not occur in the models our paper is concerned with. We may also note that no one has ever observed such a thing. 

We know detectors of the sort we describe act classically (they give out classical single clicks for photon detections, rather than superpositions of clicks). Given that we do not consider many worlds, many histories, or relational interpretations of quantum mechanics, there is no need to describe these detectors as quantum objects.

We want to emphasise that we certainly do not claim that our experiment poses a problem for quantum mechanics. It does not, as we have clearly stated in our paper. It is rather unsurprising that  \cite{Drezet2021CommentQEParadox} obtains a result compatible with quantum mechanics using quantum mechanics -- in whatever interpretation, retrocausal or otherwise.

In the local hidden variables models which we consider, the measurement outcome is always a determined detector eigenstate. To explain the experiment without the feedback loop, a non-detection at {\bf D}$_1$ must mean the photon pair was created at the lower slit, and its entangled partner had a 50:50 chance of being reflected:transmitted (and so going to {\bf U}$_3$ and {\bf U}$_4$ half of the time each). With the feedback loop, the photon would always have to go to {\bf U}$_4$. We expect this to be incompatible with observation as it contradicts the quantum mechanical result.

\textit{Acknowledgements:}
CB acknowledges support by Enterprise Ireland under the HEU award, grant number EI/CS20212057-BRACKEN.
SH acknowledges support by the Deutsche Forschungsgemeinschaft (DFG, German Research Foundation) under grant number HO 2601/8-1. JRH is supported by the University of York's EPSRC DTP grant EP/R513386/1, and the EPSRC Quantum Communications Hub (funded by the EPSRC grant EP/M013472/1).




\bibliographystyle{plainurl}
\bibliography{ref.bib}

\begin{thebibliography}{1}

\bibitem{Bracken2021QEParadox}
Colm Bracken, Jonte~R Hance, and Sabine Hossenfelder.
\newblock The quantum eraser paradox.
\newblock {\em arXiv preprint arXiv:2111.09347}, 2021.
\newblock URL: \url{https://arxiv.org/abs/2111.09347}.

\bibitem{Drezet2021CommentQEParadox}
Aur\'elien Drezet.
\newblock The quantum eraser non-paradox: a comment on arxiv:2111.09347v1.
\newblock {\em arXiv preprint arXiv:2111.13357}, 2021.
\newblock URL: \url{https://arxiv.org/abs/2111.13357}.

\end{thebibliography}

\end{document}